# Reversibility, Water-Mediated Switching, and Directed Cell Dynamics

J. C. Phillips

Dept. of Physics and Astronomy, Rutgers University, Piscataway, N. J., 08854-8019

## Abstract

**Reversible switching of the complex network dynamics of proteins is mimicked in selected network glasses and compacted small carbohydrate molecules. Protein transitions occur on long time scales ~ μs -ms, evocative of the exponentially large viscosities found in glass-forming supercooled liquids just above the glass transition; in searching for mechanisms for reversibly slowed "geared activation", Kauzmann was led from proteins to glasses [1]. I show here that selected network glasses and small carbohydrate molecules can be used to model such transitions, and elucidate in particular some universal aspects of tandem repeats. The human ankyrin tandem repeat D34, with a superhelical "coiled spring" structure [2] which has 426 residues, folds reversibly and plastically. Such molecules are too large for present transition-state numerical simulations, currently limited to ~ 100 residues solvated by ~ 3000 water molecules for times ~ ns [3]. The transition states of D34 exhibit a surprisingly simple collective ("geared') pattern when studied by fluorescence near its center, in samples modified mutageneously along its 12 helical repeats. One can understand this "plastic" pattern by taking advantage of a symmetric 45-atom carbohydrate molecular bridge to "cross over" from proteins to inorganic network glasses. There one easily identifies gears, and can show that the collective pattern is the signature of nonlocal, water-mediated [hydro(phobic/philic)] switching. Details of the transition patterns emerge from analyzing the amino acid α helical repeat sequences with water-only hydrophobicity scales. Freezing and melting of monolayer water films at physiological temperatures can enable ankyrin repeats to direct cell dynamics in muscles, membranes and cytoskeletons.**

Reversibility is fundamental to thermodynamics and statistical mechanics; far from equilibrium it has been discussed in the context of general theorems, such as the fluctuation-dissipation theorem, which has recently been applied to analyze the properties of biomolecular motors like kinesin [4]. General theorems help us to understand cyclical steady-state metabolic kinetics, but they do not differentiate structurally between systems that can achieve cyclical steady states and more common non-cyclical irreversibility. To do so one can recognize that biomolecular materials are functionally describable as complex, *evolutionarily selected*, non-equilibrium organic networks. Here we will use recent advances in the understanding of the relations between structure and dynamical reversibility of the simplest and most successful non-equilibrium networks, inorganic networks such as window glass [5], to study protein switching. Our analysis does not rely on number crunching *in silico*. The present approach relies instead on exploiting the very large and still growing data bases on inorganic network glasses as well as carbohydrates. Interpretation of these network, non-protein data bases requires the same kind of parallel analyses *ex silico* of structure and chemical bonding that eventually led Pauling to discover the α helix. Now the hope is that the combination of much larger modern data bases with the broadest theories of water-mediated interactions will lead to new insights into the molecular mechanisms of protein evolution and functionality.

**Lagrangian Constraints: Structure of Ideal Network Glasses**

It was long believed that inorganic glass transitions, like polymer transitions, are always irreversible, and that the degree of nonreversibility varies slowly with composition. However, recent experiments have identified a (nearly) reversible window (RW) in the phase diagrams of network glasses (both chalcogenide [$As_2Se_3$, $GeSe_4$] and oxide [$SiO_2$, alkali silicates and germanates, such as $(Na_2O)_{0.20}(SiO_2)_{0.80}$, and window glass, $(Na_2O)_{0.16}(CaO)_{0.10}(SiO_2)_{0.74}$]) [6], while Ag-based solid electrolytes also exhibit an RW [7]. Theory [8] and experiment [9] attribute the thermal RW to the formation of an ideally isostatic mechanical network which is rigid, but free of internal stresses.

The abruptness of the RW composition edges, and the reduction of nonreversibility within the RW, is greatly enhanced by the insertion of large, singly-bonded I anions into inorganic chalcogenide alloy networks [9], as shown in Fig. 1. This is easily understood intuitively: by reducing the connectivity and pushing chain segments apart, the repulsive lone pairs on the I anions reduce the likelihood of entanglement. By contrast, polymeric transitions are irreversibly entangled due to a high density of interchain hydrogen bonds (or alternatively, a large interchain free volume). Finally, proteins represent a nearly perfect balance between repulsive contact lone-pair forces between amino acid side groups, and hydro/(philic,phobic) water-mediated network interactions. Contact forces between amino acid side groups are topologically similar to lone pair interactions of I "dead ends" in chalcogenide alloy networks, while the effects of H bonding interactions (water-mediated network interactions) are reflected in quantitative detail in the carbohydrate data base.

Typically such isostatic (rigid, but stress-free) inorganic (chalcogenide and oxide) glass networks span composition ranges of about 10% in alloys of network formers with network modifiers. By ordering bonding forces hierarchically one finds rules for compositionally centering the isostatic RW in these prototypical networks. These rules are based on the very general idea [10] that the weaker (longer range or external) forces cause the stronger (covalent, shorter range or internal) ones to fill space optimally. For example, in Ge-As-Se chalcogenide network glasses the covalent forces, in order from strongest to weaker, are bond-stretching and bond-bending. These forces generate countable constraints $N_c$ that are reflected in bond lengths and bond angles that are preserved from crystalline compounds to glass alloys. In addition, in certain geometries, lone pair repulsive forces act as constraints. Moreover, in some oxide glasses, such as silica, oxygen bending constraints can be broken, but these constraints are restored when $N_c$ is reduced by addition of soda. Special rules have been found to describe one-fold coordinated atoms such as I (network "dead" ends). In all cases the network fills space efficiently. This leads to the quantized condition (accurate to 1% in predicting RW centers in network glasses, with no adjustable parameters)

$$N_c = N_d \qquad (1)$$

where $N_c$ is the number of intact constraints per atom, and $N_d = d$ is the number of degrees of configurational freedom with $d = 3$.

These rules have enjoyed wide successes in network glasses; they describe not only the average positions of the RW, but also explain their boundaries. This suggests that they may also provide useful guidance in understanding functions of protein networks, such as switching, where reversibility plays a key role, and kinetic processes take place slowly because of steric hindrance. To implement the latter quantitatively is very difficult in most geometries, especially in (collapsed) globular proteins, where the folding problem has been solved only partially. Because (1) works so well in the simple inorganic network context, one can hope that Kauzmann's gearing ideas can be realized explicitly in open, non-globular (network) proteins.

**Lagrangian Constraints: H Bonding and the Structure of Ideal Molecular Glasses**

Many molecular glasses are based on organic molecules where hydrogen bonding is important, and makes the molecules "sticky". (Of course, the most notable example is water, whose viscosity is large because of incipient formation of a hydrogen bond network.) Hydrogen bonding is not easily treated by any method, because the energies associated with single hydrogen bonds are small compared to covalent energies, and they become critical (compared, for example, to lone pair or Van der Waals energies) only when a hydrogen bond network involving resonating rings is formed at low temperatures (comparable to physiological temperatures). When such a network is formed around a small hydrophobic molecule, it is sometimes called an "iceberg" [11]. More generally, for large molecules with surfaces that are partly hydrophilic and partly hydrophobic, icy patches may form over hydrophobic regions.

Good, simple examples of how the ideas of covalent constraint theory taken from ideal bulk glasses can work in the context of molecular networks (where there are only a few atoms/molecule) are provided by chemical trends in the effect of H bonding on glass transition temperatures $T_g$ and other properties of supercooled liquids (such as fragility) in simple carbohydrate alcohols and disaccharides [11,12]. As shown in Table I, there are two kinds of molecular forces, the strong internal covalent forces of the carbohydrate backbones, and the much weaker external (intermolecular) H bonding forces. Little is known about the details of the latter (so far molecular dynamics simulations have "seen" these weak interactions only as "flickering" on MDS ns time scales [3], with effective temperatures ~ 300 C (hence no rings), whether on large proteins or even small molecules [13]). The two examples shown in Table I are antipodal extrema of $T_g$ in each class; both are "ideal" in the sense that proper counting of strong internal covalent *plus* weak external H-bonding constraints produces a total number of constraints which is three times the number of atoms *minus* the number of gearing coordinates, thus exhausting the available configuration space without introducing internal stresses.

The ideal H-bonded alcohol, glycerol, has the lowest $T_g$ of hydrophilic alcohols. Bioprotective trehalose, a 45 atom hydrophobic disaccharide, is an ideal H-bonded network with two molecular internal gearing coordinates that has much the highest $T_g$ of saccharides [12], and undergoes highly reversible structural transitions as part of its bioprotective anhydrobiotic functionality [12]. From these two extremes one can draw two general conclusions. First, hydrophilic interactions between covalent internal networks and water produce flexible interfaces, whereas hydrophobic interactions produce rigid interfaces. Second, counting constraints, both strong and weak, establishes the functional importance of intermolecular H bonding without specifying in detail the external geometry of how it actually occurs. This is an important point: the geometrical details are unknown, but they turn out to be of only secondary importance. In view of its many remarkable properties, it is possible that trehalose, together with the other carbohydrate results, will provide a platform for decoding the dynamics of large proteins containing thousands of atoms, much as other codes were broken using what cryptographers call keys – the Rosetta stone being the most famous example.

Trehalose has two flaps (Fig. 2), connected by a bridging O with soft dihedral angle flexibility (two "gears"). The ideal glass condition corresponds to G = 0 in the relation

$$N_c + G = 3N_A = N_d \qquad (2)$$

where G is the (integral) number of (quantum) gearing coordinates. In glycerol, where $T_g$ is extremely low, G = 0, because glycerol molecules pack poorly. In trehalose G = 2 (Table I). The two gearing coordinates correspond to the two soft dihedral angles shown in Fig. 2. In general, collective interactions between ice-like monolayers and successive trehalose flaps would not form because of accumulation of misfit stress between the ice network bound to a $C_5O$ ring structure of a single trehalose flap and a similar network bound to an adjacent flap (the diameter of the hexagonal ice ring is very nearly twice that of the $C_5O$ ring). However, from the known bioprotective properties of trehalose, it appears that the presence of two gearing coordinates is sufficient to stabilize an ice-like network that is exceptionally stable.

Arrays of trehalose molecules can form a tandem repeat structure, which identifies a (previously unknown [12]) mechanism that explains both the large value of $T_g$ and its ability to form bioprotective thin films. The tandem repeat structure is illustrated in Fig. 6 of [12a], but it can also be described in abstract terms that are equally applicable to repeat proteins. Looking at Fig. 2 here, we see that each trehalose flap has a C-O-H group outside its basic ring [$C_{11}O_{12}$ and $C_{12}O_{13}$]. In the figure these two "hooks" are on opposite sides of the two rings which can be labeled A (left) and B(right). If the entire structure is turned over, it could be labeled A*(left) and B*(right). A tandem structure is formed by the paired (double layer) series …A(BA*)(B*A)(BA*)B*… and so on. By displacing each molecule in one layer so that its outside C-O-H group alternately faces toward or away from a similar C-O-H group from the next (or preceding) molecule in the opposite layer, one builds a tandem structure of paired V's and staggered inverted V's in which the van der Waals interactions between paired rings are alternately strong and weak (Fig. 6(b) of [12a]). More generally, A and B could be any two similar but complementary structures, for instance in the protein case peptide chains with alternating positive C and negative N termini, or they could be ice-like and water-like monolayers on such chains. (Unlike covalent interactions, the ice-like monolayers are reversibly switchable at physiological temperatures, as we shall now see.) In any case, the tandem construction can be exceptionally stable, as we shall see.

**Infrared and Neutron Spectra of Trehalose and Other Disaccharides: H Bonded Submonolayers**

The mid-infrared evidence for ice-like monolayers is particularly strong for hygroscopic disaccharides, presumably because of their simple ring structures. The dry absorptance bands associated with internal covalent vibrations are broadened by intermolecular interactions. These bands **narrow and strengthen** as water is added, indicating that the water is forming a hydrophobic H-bonded network with the covalent disaccharide substrate. The changes are smallest for trehalose, presumably because of its stable, symmetrically alternating tandem repeat structure, which is present even in the absence of water. The monolayer ice (bound water) transition occurs around 343 K [14]. The spectral enhancement is maximized at a relative humidity near 50%, which is another indication that spectral enhancement is associated with cushioning of intermolecular interactions at carbohydrate interfaces by insertion of fractional water monolayers. Raman and neutron scattering studies of the lysozyme "boson peak" (which measures medium-range order) show that in the absence of water, the similarity of protein and organic solvent dynamics (organic solvent = either glycerol or trehalose) persists over 100K-350K even on ps time scales without a phase transition [14]. The atomic-scale bound monolayer ice-water disaccharide transition is probably different from the glass transition of free water confined to macroscopic protein-saccharide interfaces, measured by infrared and dielectric spectroscopy to occur around 200K [13].

**Coexisting Networks**

Constraint theory has provided an excellent, parameter-free theory of self-organized network glasses, and it is also successful in treating simple alcohols and saccharides. However, the larger question of treating hydro(phobic/philic) H-bonding of water in the presence of proteins is more complex. The problem is that both the energy and length scales of the covalently bonded protein and the ice-like H-bonded hydrophobic networks are very different; this makes it difficult to treat iceberg effects by MDS, which sees only "flickering" H-bonds, as on ns time scales ice films (interlocking $H_2O$ rings) cannot form.

In practice the problem of ice film network formation is often minimized: the interactions of amino acid side groups of proteins with water are "renormalized", and absorbed into already somewhat arbitrary classical force-field parameters supplemented by continuum solvation models [15]. The classical force field contains two parts, the first depending on point charges alone, and the second on other terms (for instance, Lennard-Jones). The two hydrophobic terms are calibrated to fit solvation energies in water and cyclohexane, which is a buckled, low polarizability and rigid $(CH_2)_6$ ring. (It would be interesting to test these continuum models against the same simple glass-forming alcohols and saccharides discussed in [12], where popular energy landscape models failed to identify trehalose interaction mechanisms [13].) This procedure is expedient for mechanical problems involving no phase changes, for example, NMR measurements of globular protein rigidity [16]. However, for transition state studies of repeats where structural changes are induced by varying the acidity of an aqueous solution, one expects ice films to melt partially (a phase transition occurs in the part of the ice-like H-bonded network), renormalization is not feasible and continuum solvation models may not be ideal. More generally one expects such partial phase changes to occur for all chemically driven protein functions. This description agrees with studies of crystallographic disorder of globular proteins which show that while hydrophobic side-chain interactions are primarily involved in determining stability of the folded conformation, hydrogen bonding and similar polar interactions are primarily involved in conformational and interaction specificity [16].

The problem of coexisting or "wrapped" networks also occurs in more complex inorganic networks; elsewhere it has been discussed [12] in the context of "double percolation", a wreath-like effect found in polymer blends, multinary chalcogenide and oxide alloys based on different building blocks (tetrahedra and trigonal pyramids), and recent quantum mechanical simulations of band-edge tail (Urbach) tail states. The knowledge base for these cases is broad and detailed, and a common pattern emerges, justifying the wreath analogy originally introduced for carbon-stained polymer blends, and emphasizing the importance of coexisting, interwoven networks.

**Minimization of Interfacial Strain Energies**

The external H bonds needed to satisfy

$$N_c + G = N_d \qquad (3)$$

produce a difference of G = 2 between $N_c$ and $3N_A$, where G represents the number of "gearing" coordinates $\varphi$ and $\varphi'$ in Fig. 2. This led to the suggestion that the remarkable bioprotective properties of trehalose, as well as its very high $T_g$, can be understood as a result of the formation of an alternating tandem repeat (hard/soft switchable) layer structure [12]. Moreover, there is strong evidence that trehalose interacts strongly (presumably through hydrogen bond networks) with not only water but also membrane lipids as films at the fraction of a monolayer level [12]. As we shall see, although trehalose is merely a very simple carbohydrate, collectively smooth H bonding [ice-like monolayers of hydrogen bonds] also explains the surprising features of tandem repeat folding in D34. In protein interfaces α helices self-organize in homodimers with parallel packing, while in heterodimers the packing is perpendicular [12].

**Constraint Theory and Globular Proteins**

The globular morphology of most proteins is determined by hydrophobic collapse, which can be utilized to construct a scale for amino acid hydrophobicity in terms of large-scale solvent accessibility [17]. However, because of the large number of atoms and the unknown role of hydrogen bonding, simple counting (as in Table I) of constraints has so far not succeeded in analyzing the transition states of globular proteins. The main features of protein structures – rigid α helices, soft loops, and super-rigid β sheets – are obviously mechanical in nature. Moreover, sequence homologies are best preserved in the more rigid elements. These factors are used as keys for analyzing many aspects of protein functionality, such as design of inhibitors [18] and analysis of ligand docking [18].

**Tandem Repeat Proteins**

D34 belongs to a broad class of tandem arrays of r ~ 33 residues (called ankyrin repeats) that form fans characterized by short-range interactions between nearly (anti-)parallel α helices with large sequence homologies [19], which function as coiled coil nanosprings, especially in membranes. D34 consists of 12 such repeats, forming two groups of n = 6 each; the entire cytoskeleton protein ankyrin contains 24 repeats, and the P(n)-distribution of (fragmentary) tandem repeats in the protein data base is approximately cut off at n = 6 repeats [19]. This cut-off can be explained by assuming that the combined effects of all the weak forces – H bonding and Van der Waals – are nearly balanced for α helices with r = 33 and n = 6. Broadly speaking, these values of r and n determine the gearing ratios through weak interactions with protein environments with distributed length scales. All processes are slowed as they search for the right sequence of gearing combinations in configuration space.

The two flaps of trehalose resemble the two groups of 6-repeats in D34, as we shall see. These tandem helical arrays are qualitatively different from the more familiar globular proteins, as they are compacted by hydrophobic forces only between repeats, leaving large areas of exposed surface normal to the local planar tangents, which could lead to large *transverse* geared interactions with, and binding to, other proteins such as actin, mediated by aqueous H-bond arrays, quite possibly not involving "noisy" interhelical loops.

Increasing urea concentrations induce protein unfolding in salt solutions that inhibit aggregation; it seems likely that urea attacks iceberg patches first, and that melting of these patches in turn destabilizes the covalent protein backbone, probably by separating regions (breaking "links") between adjacent helices [20]. (The molecular structure of urea, $((NH_2)_2CO)$, is suggestive of a wedge, with the hydrophilic O head sticking out into the water surrounding the protein, and the hydrophobic feet disrupting H bonds in the monolayer film around the protein.) The transition pattern of D34, as monitored by centered tryptophan fluorescence [21] at the nonlinearly strained 6-6 hinge [22], and probed at 7 VA mutagenic residues [23] nearly uniformly placed along the D34 array (Fig. 3), is quite unexpected [2]. Mutagenic replacement of hydrophobic [17] valine residues V by hydroneutral [17] alanine A (the most abundant amino acid) cuts off the side-chain interactions, while not affecting more distant main chain conformations [23]. For the reader's convenience Fig. 4 shows the amino acid sequence of D34 in an easily legible form, with its conserved positions, helical sections and mutagenic VA sites.

In small repeat proteins VA mutants may exhibit a weaker, but otherwise unchanged, conventional 2-state transition upon unfolding, as in the small (118 residues) ankyrin 4-repeat (probably fragmentary) protein myotrophin [24]. However, the unfolding of the much larger protein fragment D34 (426 residues) proceeds quite differently. Mutations in the 6-repeat N-terminal half of the chain broaden but do not shift the urea concentration of the transition state, while mutations in the 6-repeat C-terminal half of the chain shift the urea concentration of the transition state, but do not broaden it. This suggests that there is some kind of hinge between the N- and C-terminal halves, and that mutation in the C-terminal half has shifted this hinge to the mutated site (Figs. 4, 5, and 8 of [2]). This "plastic unfolding" behavior is strikingly parallel to the smooth hard/soft segregated (switchable) H-bond structures postulated for the two trehalose flaps on the basis of constraint theory [12], and it leads to the following quite simple, but surprisingly microscopic, model.

**Constraint Theory of Repeat Transition States: Hard/Soft Alternation**

The main features of the structural transitions of D34 can be derived *ex silico* by using the Moret-Zebende hydrophobicity scale [17], based on solvent-accessible areas of the native state. Their scale differs in many important respects from other scales, as it is not connected to specific amino acid transfer energies from water to an organic solvent. Instead, at the outset they make two very important observations. First, protein networks are divided into rigid hydrophobic parts, α helices, and soft hydrophilic loops. Hydrophobicity arises because of compression by folding of the protein chain (not necessarily, however, into a globular structure). Because the lengths of water-based H-bond networks are much larger than inter-residue spacings, hydrophobicity is an intrinsically non-local property, yet it is still residue- and sequence-specific [17]. Thus they plotted the variation of the solvent-accessible surface area (SASA) of amino acids in small N fragments ($3 \leq N \leq 35$) for 5526 protein chains (structural resolution 1A), and found that for each amino acid residue R in the center of a fragment the SASA decreased as $N^{\gamma(R)}$. Of course, such power-law (self-similar) behavior is not accidental: it is characteristic of agglomeration networks, and it is also a feature of cross-linked actin filaments in bundled cytoskeleton networks [17]. The R-dependent variations in γ arise from competition between multiple factors (size, electronegativity, dipolar moments, polarizability) that determine the self-similar curvature monitored by the SASA of Voronoi polyhedra.

Note that the absolute *(dimensionless)* hydrophobicity scale γ(R) defined in this way completely and seamlessly includes all the amino acid residues, not only hydrophobic, but also moderately and strongly hydrophilic [17], as the latter still occur, together with the former, in the rigid α helices. Moreover, there is a close parallel between the

hydro(phob/phil)ic dichotomy and isostatic Intermediate Phase regions in network glasses. The latter are rigid but unstressed, like α helices, whereas floppy regions are soft and underconstrained, much like hydrophilic loops. The smooth transition between hydrophobic and hydrophilic γ(R) is similar to that found in Ge-As-Se network chalcogenide alloy glasses composed of tetrahedral and pyramidal building blocks [6]. There is excellent qualitative agreement between the orderings of γ(R) obtained in this way, and more traditional hydrophobic orderings obtained using atomic solvation parameters derived from changes in solvation free energies in water relative to different solvents such as cyclohexane. At the same time, it is important to realize that the Moret-Zebende hydrophobicity scale is not only free of assumptions concerning other solvents (especially localization of hydrophobicity at a single residue), but it is also derived from 5526 high-resolution (1A) protein fragments, and therefore quite possibly achieves a level of accuracy far beyond what one might suppose from the reported scatter in each value γ(R). (In other words, much of this small scatter is derived either intrinsically from the 1A limits of experimental resolution, or extrinsically from distortions associated with globular compaction, beta sheets and barrels, etc., which are quite probably irrelevant to simple α repeat helices, and merely produce symmetrical background.) One can hope that the Moret-Zebende hydrophobicity scale includes water-mediated evolutionary correlations implicitly, in accord with theories of exponentially complex molecular evolution [17]. Thus it will enable us to estimate hydrophobic repeat effects without knowing where the water is, just as constraint theory succeeded in treating the dimensionless gearing effects of external H-bond interactions of alcohols and saccharides without specifying those interactions in detail. For this purpose, D34 is an ideal first test case.

Given this background, one can proceed to analyze D34 in the simplest possible *ex silico* way. Each repeat contains two α helical "chopstick" segments that form an L (residues 4-11 and 15-23, hereafter designated as $\alpha_A$ and $\alpha_B$, respectively; note that these two "arms" have nearly equal lengths, which enhances their resemblance to the two flaps of trehalose). All the S = 13-24 repeats form a curved hook or coiled spring structure, and each repeat is formed from two curved helical "chopstick" segments A (convex, outer) and B (concave,inner) (very pretty: fractal α helical Riemannian curvature embedded in a higher S fractal repeat Riemannian space, rather like Ezikiel's "wheel in a wheel"). We can estimate the fractal curvatures $\gamma_A(S)$ and $\gamma_B(S)$ to first order in their deviations from the average values quite well simply by averaging $\gamma(R)$ over residues 4-11(A) and 15-23(B), respectively, for each S. (In conventional language, $\gamma_A(S)$ and $\gamma_B(S)$ are configuration coordinates.) When we do this, a surprisingly simple picture emerges: for the outer A arms, $<-\gamma_A(S)>$ is nearly constant near 0.155 (hydroneutral), while for the inner B arms $<-\gamma_B(S)>$ is hydrophobic and is nearly constant near 0.166, as is their average $\gamma(S)$ near 0.16, except for S = 18 (strongly hydrophobic, average = 0.17), and S = 23, average = 0.15 (hydrophilic). Thus S = 18 is a (hydro)phobic hinge between the N and C ends of the repeat, while S = 23 is a hydrophilically weakened switch for the entire repeat. The two repeat regions S = 13-17 and S = 19-22 are quite analogous to the two flaps of trehalose (Fig. 3), as are the N and C ends of the repeat. An attractive feature of the calculation is near compensation of the fractal helical hydrophobicity of the combined 18 and 23 repeats (nearly hydroneutral), reflecting a consistent overall fractal hydrocurvature, so that the protein is nearly perfectly balanced internally and externally.

While one might have expected that the loops would always be hydrophilic, Fig. 6 shows a soft/hard alternation of hydrophilic loops with hydrophobic pinning ones associated with binding the spectrin domain in a cleft formed by repeats 20-24. Considering that the oscillations in Fig. 6 stem entirely from the amino acid sequence, their correlation with the structural cleft is excellent evidence for the importance of water-protein interactions in evolutionary sequence selection. Returning now to Fig. 5, it appears that the three B-arm comb-like hydrophobic peaks at S = 16, 18 and 20-21 contribute to the membrane-binding functionality of ankyrin R. Finally, the overall hydro index for all the D34 amino acids is 0.153 - nearly hydroneutral.

Another way to analyze the repeat structure is to define a second configuration coordinate, the hydroflexibility index $\Phi(S)$ for each α helical segment S, in terms of the marginal stiffness induced by its water-mediated adjacent amino acid pair interactions by

$$\Phi(S) = \Sigma[(\gamma_i(R(S)) - \gamma_i(R(S+1)))^2 + (\gamma_i(R(S)) - \gamma_i(R(S-1)))^2 ]/2M \qquad (4)$$

Here R denotes an amino acid and $\gamma_i(R)$ is its hydrophobicity; the sum is over matched helices of maximum length M, so that $\Phi$ is normalized.. For repeat adjacent helical segments composed entirely of consensus residues, $\Phi = 0$. To the extent that interactions between adjacent repeats are mediated by ice-like patches, $\Phi$ can be a useful measure of repeat flexibility; for example, the results that α helices at homodimer interfaces prefer parallel packing [12] are consistent with stabilization by reduced hydroflexibility. For D34 the results shown in Fig. 7 are spectacular: $\Phi_A(18) = \Phi_B(18)$, which means that the S = 18 central repeat is not only hydrophobic, but also that the A and B arms are equally flexible, and the entire repeat can switch as a unit without distortion. The results for the S = 16 repeat are equally spectacular: The A(16) arm is nearly rigid, while the B(16) arm, probed by the V519A mutation, is very flexible. The very large flexibility of the B arm for S = 24 is presumably connected with the long, strongly hydrophobic (Fig. 7) C terminal loop to which it is attached.

To analyze transition state data obtained by denaturization with urea one should have some idea of the interactions between urea, proteins, and water. The usual view is that urea disrupts the hydrogen bond network of water and behaves much like a hydrophobic protein monomer [20]. The largest destabilizing effect of urea ($(NH_2)_2CO$) on a repeat would be to act as a hydrophobic wedge between the A and B helical arms. This model can now be compared to the fluorescence and circular dichroism data of [2], with successive VA mutations at position 18 near the center of $\alpha_B$ (15-23) in repeats 13, 15, 16, 18, 20, 22, 23, that reduce the hydrophobicity of each substituted repeat in turn by ~ 0.01. These shifts, while significant, are not large enough to alter the main features of D34, which are the rigid S = 18 central phobichinge between the N and C ends of the repeat, and the S = 23 hydrophilic switch for the entire repeat. The fluorescence signals broaden as the mutations approach the C and N ends, but even after allowing for this, the signal in the mutated strongly hydrophilic (softened) repeat 23 (V750A) is exceptionally broad (Table 1 of [2]).

It appears that V519A, in the B arm of repeat 16, is associated with three extremal features of the transition state data. The center of the urea concentration of the fluorescence transition is smallest for this mutation, as is the slope parameter $m_{I-N}$ (a broad transition, easily reached). At the same time the Circular Dichroism of the V519 mutated protein shows the largest $m_{I-N}$ value (narrow transition). Looking at Fig. 5, we see that for the N half of D34, the difference in hydrophobicity between the $\alpha_A$ and $\alpha_B$ helices, $< -\gamma_B(S)> - < -\gamma_A(S)>$, is largest in repeat $S = 16$. This immediately suggests a broad fluorescence transition, with a small center concentration (the A and B helix arms should react differently to the acid). However, CD is sensitive to helical hydrocurvature, and may well depend on the A and B hydrocurvature difference, requiring both helical arms, so that there is a sharp transition as soon as one arm unwinds. Such a transition is also suggested by the large difference between $\Phi_B(16)$ and $\Phi_A(16)$. For all these reasons, the wedging effect of urea should be largest in repeat 16, so that theory explains the extremal properties of m(16) in both the fluorescence and CD data. Finally, as interpreted in Figs. 4 and 5 of [2], D34 is partitioned near its center into N and P parts. The similarity of this partitioning to the structure of trehalose is obvious.

The present model says that *all* the transition states (probed by VA mutations of different repeats) of D34 are very likely to be associated with unfolding the A-B arms of repeat 16, as urea wedges heteronucleate between these arms. The broadened and shifted transition states associated with mutations of other repeats merely reflect the different ways that urea condensation at repeat 16 has of reacting to the weakening of the A-B interaction in each respective repeat by the VA mutation in the B arm of the relevant repeat. This explains in the simplest possible way why the transition has extremal properties when the VA mutation occurs in repeat 16, and why in that case the transition is sharper than in the wild case. It also explains why the observed fluorescence m values vary monotonically with separation from repeat 16, increasing in both directions. Because repeat 16 is in the N-terminal half of the protein, mutations there always sharpen (larger m) the transition relative to the wild type. However, because of the hydrohinge at repeat 18, in the C half the mutated transitions are broadened relative to the wild type (smaller m). In effect these mutations have a larger weakening effect on the hinge itself than they do on the A-B arms of repeat 16, thus damping the transition at the latter. This model is especially appealing in the context of the large difference in hydrofragilities of the repeat 16 A and B arms in Fig. 7, which is mainly caused by strong hydrophobicity similarities in the 15, 16 and 17 A arms.

It is instructive to return to the RW in network glasses. Generally this window spans a fractional range of covalent constraints of about 10%, but the window can be narrowed to 1% (and also deepened by a factor of 10, to be nearly reversible) by reducing constraints by replacing Se connectors with I termini in g-GeSe$_3$ [25], as shown in Fig. 1(b). The non-bridging I ions presumably function as weak contacts (van der Waals hinges). In much more sophisticated evolutionarily evolved protein networks, with many interchain weak contacts, the RW should be very narrow. Then the presence (absence) of external collective hydrogen bonding can alternately harden (soften) apparently closely similar (in trehalose, ergodically identical flaps) ankyrin α-repeat building blocks, reduce window width to nearly zero, and create nearly reversible mechanics. Here in D34 this hard/soft alternation occurs on two levels, between the A(S) and B(S) helical arms of each repeat S, and between the N and C halves of the repeat array itself. Note that the theme of alternating hard/soft nanocomposites is also found in bone [26] and the muscle protein titin [12,27]. (On the much larger genomic scale isostatic/floppy alternation appears necessary to the loop dynamics presumed to explain local and nonlocal aspects of gene functionality [28]. Floppy sections are also natural candidates for "junk" DNA (transposable elements) [29].) One can go further in ankyrin proteins, and argue that it is the nonlinear accumulation of strain energy that produces the cutoff in the P(n) repeat distribution at n = 6 [19].

Quite generally the binary hard/soft feature of both trehalose and D34 enlarges the coherence length scale. It also introduces collective coordinates that can bundle together and thus stabilize many weak external hydrogen interactions to either water or other proteins, such as cytoskeletal actin. It is natural to suppose that this mechanism is sufficient to replace the flickering individual H-bond interactions (effectively fluctuating at a temperature ~ 300C) seen in ns molecular dynamics simulations [3,13] with physiologically meaningful (ms, slow) condensed or collective H-bond modes [ice-like monolayers of hydrogen bonds], similar to those previously suggested as the origin of the bioprotective properties of trehalose [12]. This means that ankyrin repeats, for example, can not only bind to ubiquitous proteins such as actin, but can also exploit their full 3-dimensional structure (including the L-shape ("cupped hand") of each repeat [30]) to function as *binary switches* that *direct* the large-scale motion of polymerized globular proteins (such as actin) in cytoskeletons [31]. In this way living cell switches can avoid the H bond entanglement that makes polymer melting irreversible. The collective model is also consistent with studies that show that load fluctuations drive actin network growth [31].

**Other Examples**

Collective interfacial H-bond interactions probably also explain the shifts in opposite directions of the glass transition temperatures of salol (G > 0) and glycerol (G = 0) confined in Si mesopores [32]. A protein and small molecule database survey showed that H bonds are divided into two sets, strong and weak, and that both strong and weak hydrogen bonds are involved in ligand binding [33]. Mechanical criticality (proximity of a protein to a stiffness transition) is evidently a possible explanation for (1) the mechanism of rate-limiting motions in enzyme function, which can be altered by replacing histidine by alanine at a single site of 160 potential exchangeable protons in RNase [34], and (2) the slowing of folding of an intestinal fatty acid by more than three orders of magnitude by side group replacement [35].

**Ideal Neuronic Switches**

Because ankyrin repeats are attached to transcellular proteins like polymerized titin and actin, they are not fully symmetrical, as they are preferentially tethered at one end. Fully symmetrical dimers (like trehalose) are natural candidates for ideal binary superswitches. It is striking that excitatory synaptic transmission in the central nervous system is mediated by symmetric dimers (150 residues each) separated by an ion channel in glutamate receptors [36]. As in the ankyrin repeats, the rigid units of the dimers are again only α helices, while the ion channel is a natural site for collective transverse H bond ice-like monolayer interactions. The latter are a symmetrically attractive supplement to the stretched covalent longitudinal tethering model proposed in Fig. 4 of [36].

**Universality and Marginal Elastic Stability**

The key theme of this paper is that physical, organic, and biophysical glasses share many common features (universality). These are already reflected in the similar glass-forming tendencies of inorganic and organic network glasses [12]. Upon supercooling slowly, in order for a glass to form, instead of crystallizing, it should form a network that is neither too rigid, nor too flexible, in other words, it should form an ITP, or it should have marginal elastic stability, with zero internal stress [9]. It is easy to test repeat proteins to see whether their wild forms have marginal elastic stability, and in fact the test is so obvious that it has already been carried out both in NMR [37] and in AFM [27] experiments. One identifies the "consensus" amino acid α helix repeat sequence, and compares the stability of synthetic proteins composed of consensus repeats with wild mixed repeats. According to the universality hypothesis, the wild forms have evolved to have marginal elastic stability, so that the synthetic consensus repeats should be more stable, and they are [37,27]. Apart from the ankyrin repeats, the superhelical HEAT repeats are noted for their inherent conformational flexibility [38].

From the point of view of protein science, the enhanced stability of synthetic proteins composed of consensus repeats is quite obvious, as it is well known that amino acid repeat sequences are always better conserved in stiff structural elements (α helices) than in flexible ones (loops). However, there is a much more exacting test for network universality, which is still regarded as controversial. The electronic phenomenon of high temperature superconductivity (HTSC) depends for high transition temperatures on strong electron-phonon interactions. These electron-atom interactions in turn screen atom-atom interactions, destabilizing the lattice against Jahn-Teller distortions: high $T_c$'s are correlated with soft structures. The phenomenology of composition-dependent least-upper-bound $T_c$'s suggests that marginal elastic stability accurately dominates all other factors and is the key factor determining chemical trends in the very well-studied cuprate HTSC [39].

**Globular Proteins**

While the foregoing model fulfills many of the hopes voiced earlier [12] regarding the value of relating the geared slow dynamics of "transition states of [proteins to] the reversible-non-aging window of network glasses", it also contains many cautionary features. The repeat structure of D34 ankyrin proteins is nearly symmetrical and exceptionally simple, as are the glutamate symmetric dimers, which is why the trehalose analogue analysis used here may be relevant. One can recognize only a few similarities in the folding dynamics of globular proteins, with their rigid elements packed densely together to form extremely complex structures and correspondingly complex folding pathways [3,13]. A relatively simple globular case is the 452-residue beta-barrel enzyme fragment *ePRAI*, whose H-bonded surface seems to destabilize before its interior [40]. This surface/volume effect might be characteristic of single-domain globular proteins [40], although it is not obvious from a coarse-graining analysis of the rigidity of the covalent backbone alone [41]. Moreover, even wild globular proteins are less stable against mutations than synthetic ones, in other words, marginal elastic stability appears to be a universal feature of both globular and repeat proteins [40].

An enzyme that is much closer to D34 is LacY, the well-studied lactose permease of *Escherichia coli*, which also contains 12 α helices in 6-6 dimers that function as switches for proton and disaccharide transport across membranes; however, unlike D34, the sequence homology is low, even between the two dimers [42]. This could reflect mismatch between the hydrophobic surface of the protein and the hydrophobic interior of the lipid bilayer. The asymmetry of the dimers could induce flexing or bending of a transmembrane segment of the protein [43]. In any case, LacY illustrates the suitability of disaccharide analogies, based on trehalose, as a basis for identifying the functionality of repeats as biophysical switches with a very wide range of contexts (even when homologous repeats are not obviously involved).

**Conclusions**

This paper has examined three exponentially complex, non-Newtonian problems associated with space-filling, compacted glassy networks: prototypical network glasses, small hydrophilic alcohols and hydrophobic saccharides, and repeat proteins. Traditional Newtonian methods, based on classical force fields and "brute-force" polynomial number crunching, have achieved modest successes on these problems. However, Lagrangian theories have led to the discovery of the reversibility window in network glasses (overlooked for decades using traditional methods), as well as simple yet powerful characterizations of the glass-forming tendencies of alcohols and bioprotective properties of saccharides, especially trehalose. This paper has shown how similar non-traditional methods can be used successfully to understand protein-water interactions with exceptional amino-acid specificity.

Relatively small proteins with fewer than 150-200 residues exhibit simple structures, with residue depths dominated by surface effects [44], and changes from the native state to the transition state in Solvent Accessible Surface Area (ΔSASA) that are linear in the number of residues [45]. Above the ~ 175-residue slope crossover, structures are dominated by core interactions [46]. Repeat dynamics in 420-residue D34 occurs plastically on a large scale because the $S = 18$ repeat can function as a geared hydrophobic switch between the N and C ends, with a small hydrocurvature radius concealed in its iceberg covering. Similarly, the hydrophilic $S = 23$ repeat can function as a pivot for the entire protein. Note that the Moret-Zebende hydrophobicity scale is ideally suited to describing the 33 residue repeats as units, because the pure numbers that it has derived for its scale are based on SASA for the amino acid helical length range between 3 and 35. These pure numbers are the natural Lagrangian generalizations of constraint theory for inorganic glass networks and H bonding constraints for small alcohol and saccharide molecules; however, they go much further, because their nonlocal aspects incorporate the self-similar curvatures that have developed through evolution.

While it may be too early to discuss global trends, studies of several other repeat proteins (discussed elsewhere) have shown hydro patterns quite different from that of D34. Those patterns correlate well with functionality. It appears that the distinctive property of D34 is that its A arms are always more hydrophilic (stabilized to the outside) than the B arms (stabilized to the inside), except at the central repeat 8. Thus D34 is ideally suited for attaching to polymeric actin and titin and so functioning as a cellular switch.

It appears plausible that the switching mechanisms for glutamate neural receptor domains and ankyrin repeats bound to titin (muscles) [47] and actin (cytoskeletons [48]) are semi-quantitatively similar, and may all be based on quantized gearing driven by ice-like films. All are based on protein domains that are symmetric or nearly symmetric, and for which a simple carbohydrate (in the simplest case, "two residues and two gears"!), fully symmetrical prototype (trehalose) is known, and all could involve a geared order-disorder (ice-water) monolayer transition. Earlier it was thought that "the myriad of different roles for ANK repeat proteins makes a common function, such as [lock-and-key] enzymatic activity, extremely unlikely" [30], and [their function is] "as yet unclear" [22], but it now appears, especially from the data of [2], that hydrophobic switching is just that "unlikely" common function. Because reversible switching is basic to many biofunctions, it is easy to understand why recent genome searches have identified an amazingly large number (10,000 to 20,000) of ankyrin repeats [49].

The methods of protein network analysis developed here are quite different from currently fashionable mechanical models. On the one hand, molecular dynamic models based on classical force fields inevitably introduce local approximations to amino acid-water interactions that neglect the long-range self-similar correlations found in loops and α helices and identified by the Moret-Zebende hydrophobicity scale; in classical models the local hydrophobicity of a given residue in a loop is the same as that for the same residue in an α helix. On the other hand, quantum non-local "first principles" treatments of homogeneous model hydrophobic surface potentials are not residue-specific and still do not include long-range self-similar amino acid α helical correlations.

Here analysis has shown that the Moret-Zebende hydrophobicity scale, based on Voronoi tessellations and α helical solvent-accessible surface areas that decrease fractally with increasing helical length up to $N = 35$ residues, appears to be very accurate, and yields a satisfactory picture of the mutagenesis dependence of transition states in D34, a nearly symmetrical ankyrin repeat. Other global, water-only approaches to hydrophobicity (such as residue depth) are somewhat ambiguous [44] and depend on protein size [46]. At present there is no rigorous justification for the M-Z scale, except the intuitive one that it is unambiguous (seamless, complete and dimensionless, thus free of size effects, apart from the unavoidable cutoff at N = 35). The M-Z scale measures the rate of reduction of the SASA with helical growth, which presumably reflects evolutionary residue correlations mediated by interactions with water. The glassy analogies (especially with the IP in network glasses, and glycerol and trehalose in small organic molecular glasses) discussed here might be extended to growth of contacts in glass foams [50] to provide specific analogies for the fractal powers identified by M-Z. Just as the M-Z exponents depend on the central residue, the glass foam exponents depend on the initial state; moreover, while the latter are > 1 for initially crystalline bubble arrays, they are < 1 for disordered bubble arrays, thus resembling the M-Z fractal exponents.

Theorists interested in the reversible aspects of sequence-enabled, large-scale protein dynamics (including protein-protein interfaces) could well concentrate their attention on trehalose and these easily accessed non-globular repeat proteins, which apparently share many functional as well as structural features, conveniently including a more accessible external hydro(phob/phil)ic H bonding pattern. Because the structures of repeat proteins are so much simpler than those of globular proteins, it may well be possible to develop new large-scale methods (of which the Moret-Zebende hydrophobicity scale is a successful example) that are specific to these ideal cases. It also seems likely, regardless of the "rigor" of the M-Z scale (which has passed the D34 test with flying colors), that it can provide many further valuable insights into the nature of water-protein interactions unobtainable by traditional methods. Searches for ARM and HEAT repeat proteins at present rely mainly on identification of conserved residues at short interhelix turns [51]; this analysis could be supplemented by looking for hydro patterns in helical arm and loop sequences.

| Material | $3N_A$ | $N_c$ | $N_H$ | $N_c + N_H$ | G |
|---|---|---|---|---|---|
| glycerol | **42** | 31 | 11 | **42** | **0** |
| trehalose | **135** | 117 | 16 | **133** | **2** |

Table I. A condensed version of Table II of [12], which included other alcohols and saccharides, and broke down the H-bonding into O-H··· and C-H··· stretching and bending components. The key point of [11] is that one can use the condition $3N_A = N_c + N_H + G$ to determine the optimal sets of H bonds ··· that are consistent with the known glassy properties of glycerol and trehalose. These led to a model of trehalose as an alternating tandem repeater with G = 2 adaptive coordinates. Similar considerations may explain how small ankyrin repeats can function as switches at widely separated branching points of the cytoskeleton, thus connecting "geared" molecular and cellular morphologies.

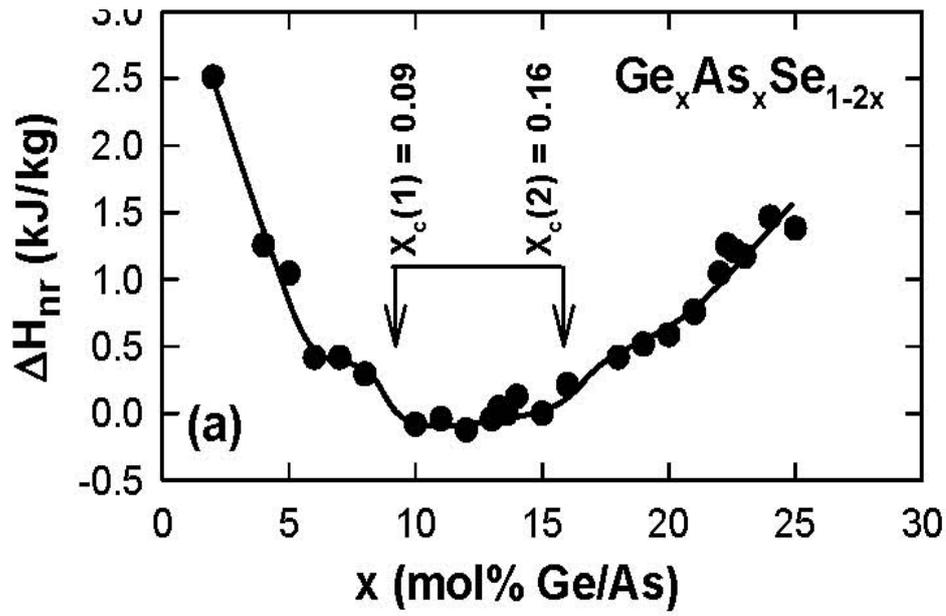
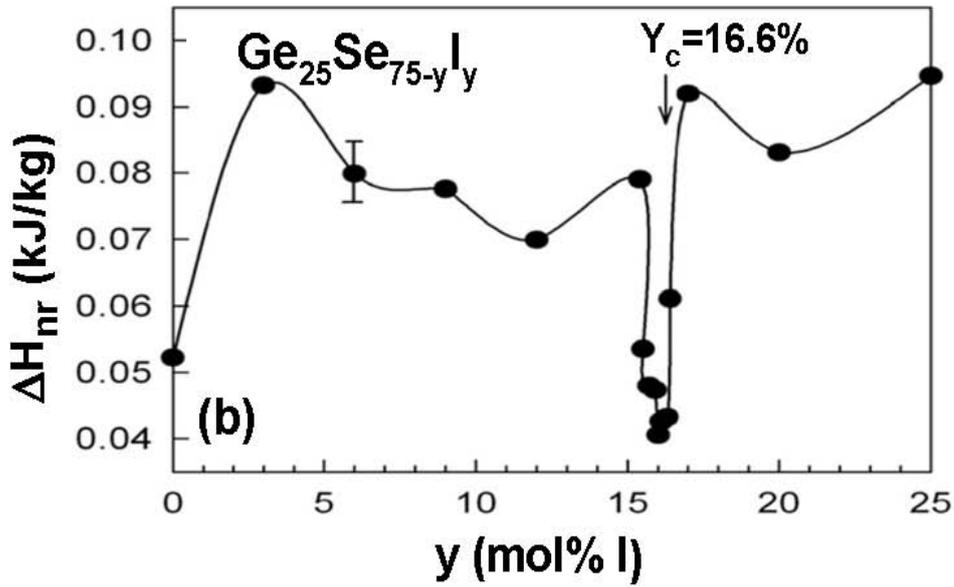
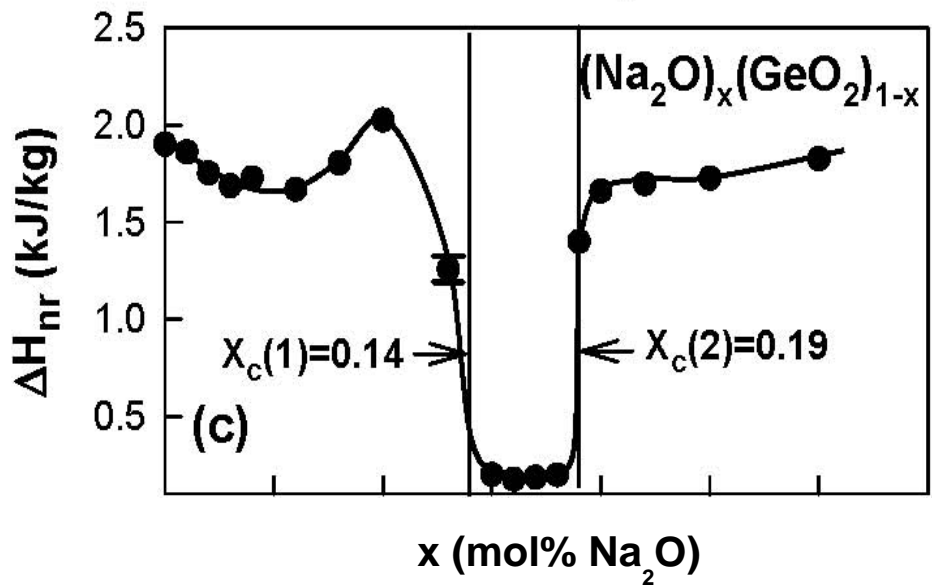

Fig. 1. Reversibility windows (RW) in network glasses. (a) Typical RW for fully cross-linked ternary chalcogenide glass alloys with isovolume elements. (b) Designed chalcogenide alloy RW utilizing I ions to reduce cross-linking and narrow RW. (c) Oxide RW with cross-linking reduced by Na ions.

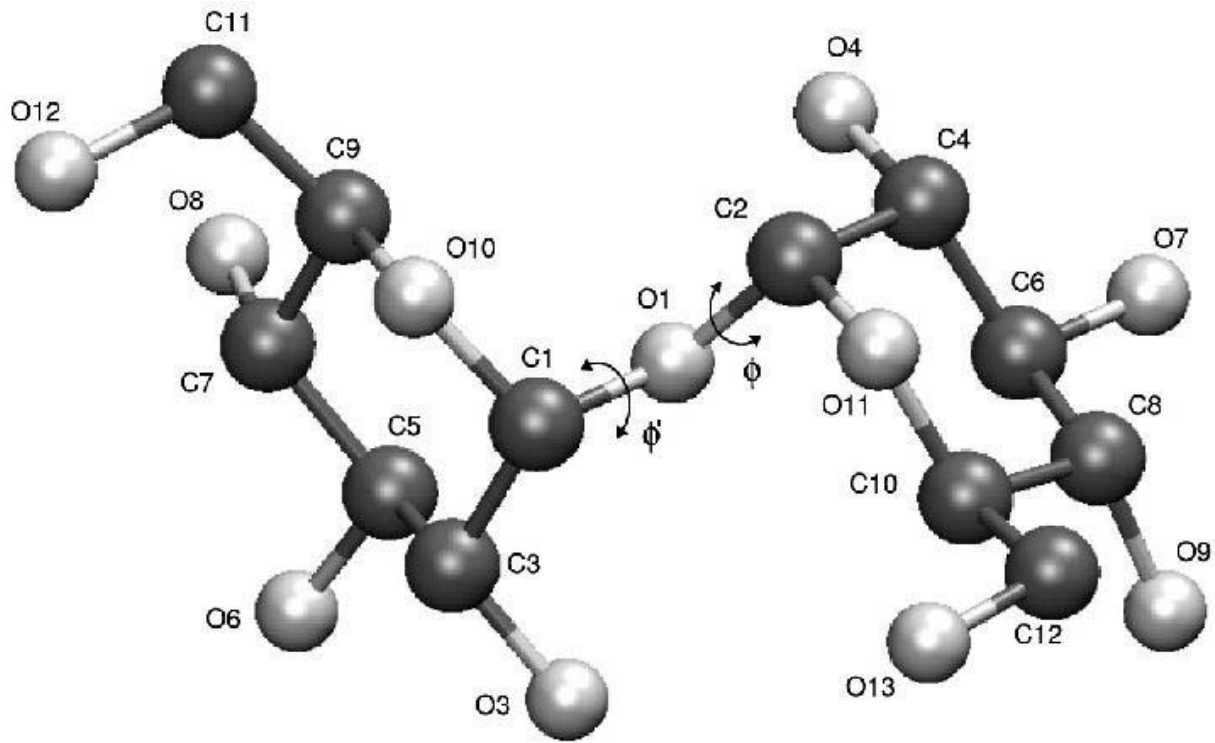

Fig. 2. The molecular structure of trehalose, with hydrogen atoms omitted [12]. The gearing coordinates are φ and φ'.

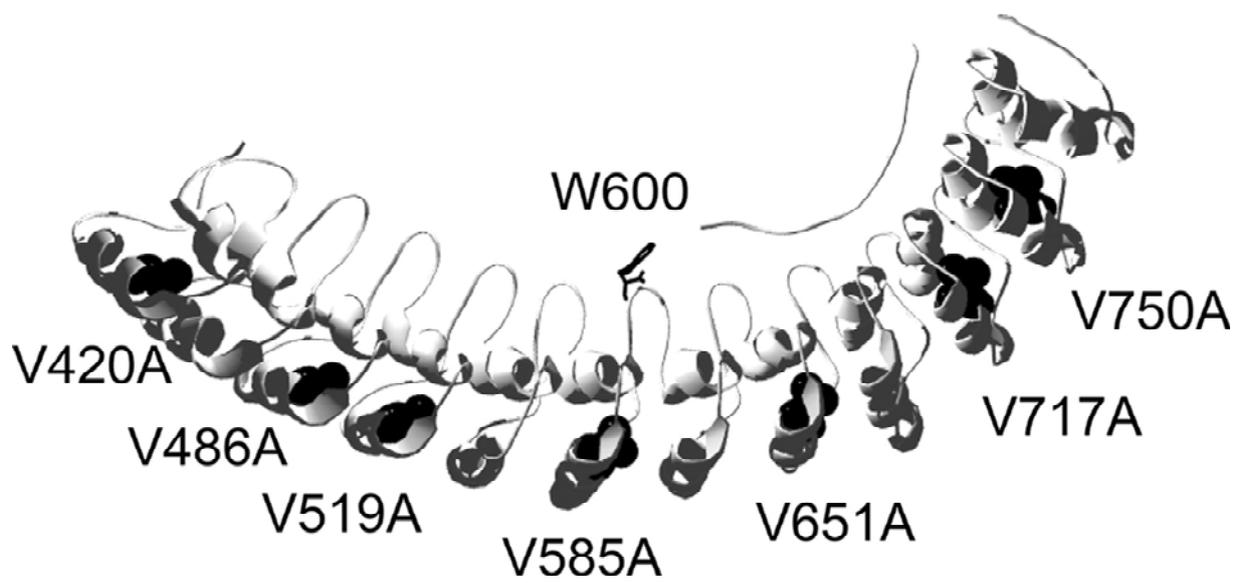

Fig. 3. The protein fragment D34 with a tryptophan fluorescent monitor centered at W [21], as mutated with transition state VA probes [23] at the marked sites [2].

13  SGL *TPLH V AS* FM **GH** *LPI V\* KN LL Q*R **GA** SPN VSNV 435

14  KVE *TPLH M AA* RA **GH** *TEVA KY LL Q*N **KA** KVN AKAK 468

15  DDQ *TPLH C AA* RI **GH** *TNM V\* KL LL* E N **NA** NPN LATT 501

16  AGH *TPLH I AA* RE **GH** *VET V\* LA LL* E K **EA** SQA CMTK 534

17  KGF *TPLH V AA* KY **GK** *VRV A EL LL* E R **DA** HPN AAGK 567

18  NGL *TPLH V AV* HH **NN** *LD I V\* KL LL* PR **GG** SPH SPAW 600

19  NGY *TPLH I AA* KQ **NQ** *VEV A RS LL Q*Y **GG** SAN AESV  633

20  QGV *TPLH L AA* QE **GH** *AEM V\* AL LL* SK **QA** NG NLGNK 666

21  SGL *TPLH L VA* QE **GH** *VPV A DV LI* KH **GV** MV DATTR 699

22  MGY *TPLH V AS* HY **GN** *IKL V\* KF LL Q*H **QA** DVN AKTK 732

23  LGY *SPLH Q AA* QQ **GH** *TDI V\* TL LL* KN **GA** SP NEVSS 765

24  DGT *TPLA I AK* RL **GY** *ISV T DV LK* VV **TD** ETS FVLV 798

S D K  H R M S F  PE  TV  DE  I L D  V  SE   DE  GE  EL   I S F

Fig. 4. The amino acid sequence of D34 [19], with row numbers S = 13-24. The consensus repeats are shown in bold letters in *columns*, the α helices in red italics, and the mutagenic VA sites [2] are marked by asterisks: note that these always occur near the center of the B helical arm. This 420 residue sequence can be decoded *ex silico* by using the Moret-Zebende hydrophobicity scale to identify chemical trends in the repeat *rows* (see text and Figs. 5-7).

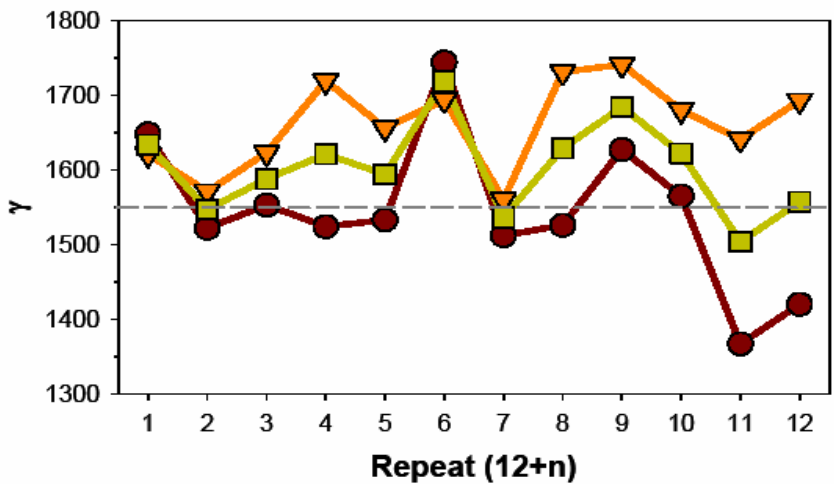

Fig. 5. Average <-γ>'s for the A (4-11) helical segments (brown circles), and B(15-23) helical segments (orange triangles), of D34 amino acid helical repeat sidegroups listed in Fig. 4, and their average (green squares). The dashed line represents hydroneutrality (-γ = 0.155; the decimal points are omitted from the γ coordinates). The differences between the A and B helices are smaller near the N end and at repeat 18, which is an extended hydrophobic hinge. Near the C end the B helical arms, which are attached to longer loops, remain hydrophobic, while the A helical arms become strongly hydrophilic, and make repeat 23 an extended hydrophilic hinge. The qualitative hydrophobicity differences between the N and C ends are consistent with the opposing trends (quenching in the C end mutations, enhancement in the N end) in fluorescence seen in [2]. Further details are discussed in text.

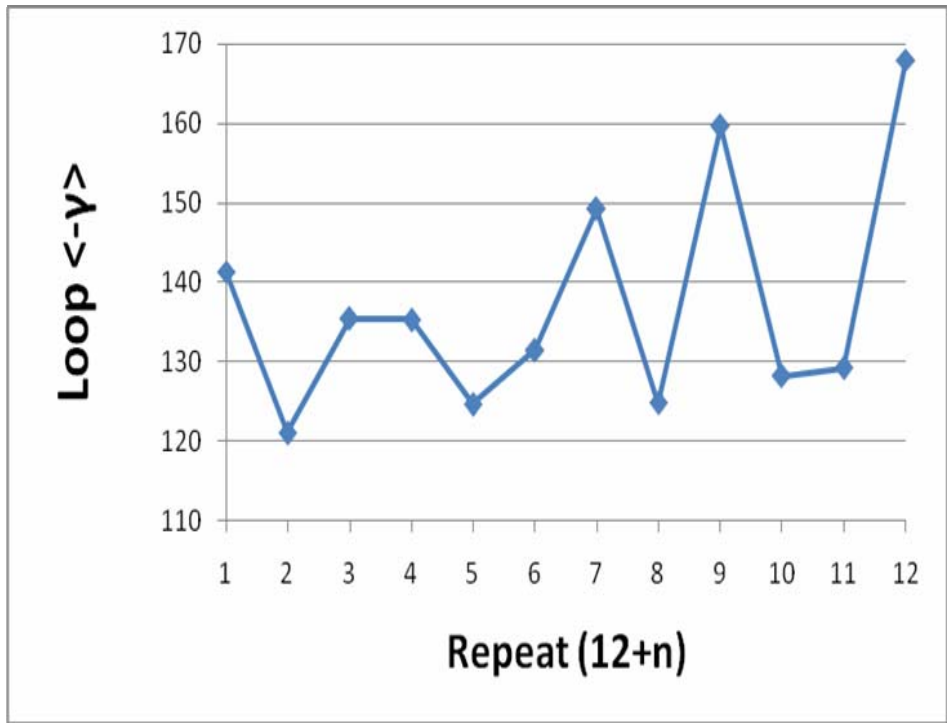

Fig. 6. Most of the loops are hydrophilic (<-γ> = 0.13, or 130 in ordinate units here, relative to hydroneutral = 155), but there are exceptions. The S = 24 (n = 12) loop, which connects the C terminus directly to the spectrin-binding domain, is strongly hydrophobic, as (decreasingly) are the S = 21 and S= 19 loops, which also bind the spectrin domain in a cleft formed by repeats 20-24 [37]. This hydrocomb structure is echoed in the hydrofragility pattern shown in Fig. 6.

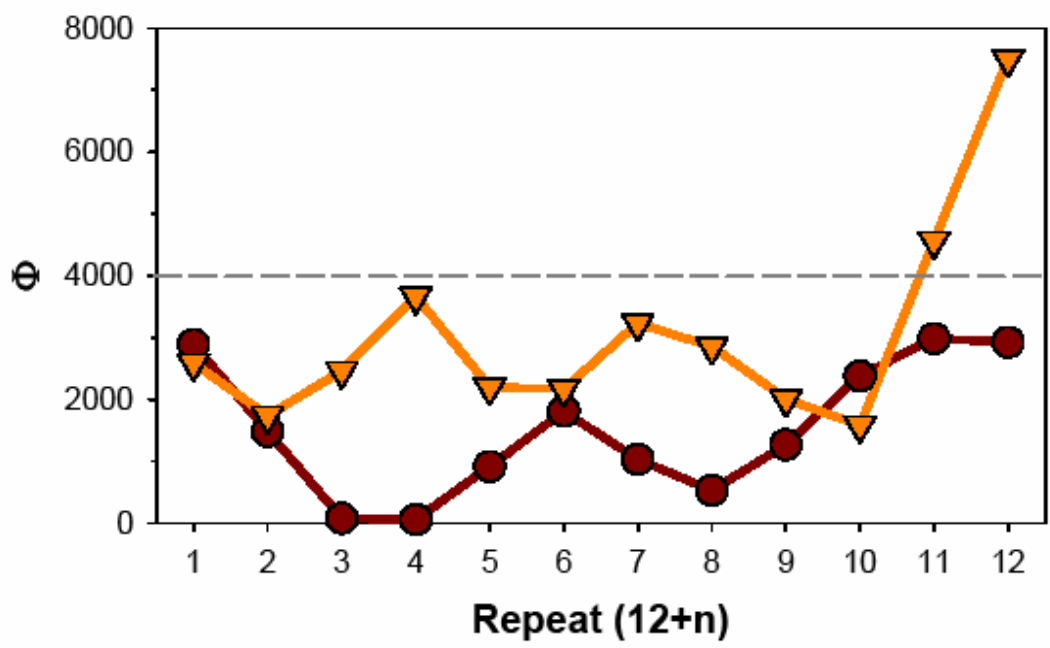

Fig. 7. Average hydroflexibilities $\Phi_A(S)$ and $\Phi_B(S)$ for the A (4-11) helical segments (brown circles), and B(15-23) helical segments (orange triangles), of D34 amino acid helical repeat sidegroups listed in Fig. 4. The A arms have many consensus repeats and are generally more rigid than the B arms, which become very flexible as they approach the long loop attached to the C terminal. The large differences $\Phi_B(16) - \Phi_A(16)$ and $<-\gamma_B(16)> - <-\gamma_A(16)>$ probably determine the exceptional sharpening of the fluorescent and CD transitions in V519A mutated D34, as S = 16 (n = 4) not only has a B hinge (local maxmum in $\Phi$) with a rigid A arm, but also the B arm is a local maximum in hydrophobicity, while the A arm is hydrophilic (Fig. 5).

.